\documentclass{article}
\usepackage{subcaption}
\usepackage{arxiv}
\usepackage{enumitem}
\usepackage[utf8]{inputenc}
\usepackage[T1]{fontenc}    
\usepackage{hyperref}       
\usepackage{url}            
\usepackage{booktabs}      
\usepackage{amsmath}   
\usepackage{nicefrac}       
\usepackage{microtype}  
\usepackage{lipsum}		
\usepackage{graphicx}
\usepackage[numbers,sort&compress]{natbib}
\usepackage[scr=rsfso,frak=euler,bb=ams]{mathalfa}
\usepackage{doi}
\usepackage{xargs}
\usepackage[pdftex,dvipsnames]{xcolor}  
\usepackage{algorithm}
\usepackage{graphicx}%
\usepackage{multirow}%
\usepackage{mathrsfs}%
\usepackage[title]{appendix}%
\usepackage{xcolor}%
\usepackage{textcomp}%
\usepackage{manyfoot}%
\usepackage{booktabs}%
\usepackage{algpseudocode}%
\usepackage{listings}%
\usepackage{slashbox}%
\usepackage[colorinlistoftodos,prependcaption,textsize=tiny]{todonotes}
\newcommandx{\comment}[2][1=]{\todo[linecolor=red,backgroundcolor=red!25,bordercolor=red,#1]{#2}}

\def\bra#1{\mathinner{\langle{#1}|}}
\def\ket#1{\mathinner{|{#1}\rangle}}

\newcommand{\ketbra}[2]{|#1\rangle\langle#2|}

\title{Quantum Agents for Algorithmic Discovery}
\date{}

\author{ 
	\hspace{1mm}Iordanis Kerenidis \\
    Quantum Signals, Paris, France \\
    IRIF-CNRS, Univ Paris-Cité,
	Paris, France \\
	\texttt{iordanis@quantumsignals.ai} \\
    \And
    \hspace{1mm}El-Amine Cherrat \\
    Quantum Signals, Paris, France \\
    \texttt{amine@quantumsignals.ai} \\
}
\renewcommand{\shorttitle}{Quantum Agents for Algorithmic Discovery}

\begin{document}
\maketitle

\begin{abstract}
	We introduce quantum agents trained by episodic, reward-based reinforcement learning to autonomously rediscover several seminal quantum algorithms and protocols. In particular, our agents learn: efficient logarithmic-depth quantum circuits for the Quantum Fourier Transform; Grover's search algorithm; optimal cheating strategies for strong coin flipping; and optimal winning strategies for the CHSH and other nonlocal games. The agents achieve these results directly through interaction, without prior access to known optimal solutions. This demonstrates the potential of quantum intelligence as a tool for algorithmic discovery, opening the way for the automated design of novel quantum algorithms and protocols.
	
\end{abstract}
\section{Introduction}\label{sec1}

Quantum computing has redefined computation by leveraging principles fundamentally distinct from those governing classical computers. Unlike classical computers, which process information using bits, quantum computers use qubits. These exploit superposition and entanglement, enabling fundamentally new modes of computation. This paradigm shift has led to the development of a number of quantum algorithms, such as Shor's factoring \cite{shor1994} and Grover's search \cite{grover1996} algorithms, which have demonstrated the potential to solve certain problems provably faster than their classical counterparts. Quantum cryptography has likewise introduced cryptographic primitives, such as quantum key distribution \cite{bennett1984}, offering information-theoretic security under standard assumptions.

Despite these promising advances, the discovery of new quantum algorithms and practical quantum applications remains a great challenge. The intricacies of quantum information and the absence of large-scale quantum computers make the design, implementation, and benchmarking of novel quantum algorithms a complex task. Furthermore, translating the theoretically provable quantum advantages into real-world applications is fraught with difficulties, in particular due to the far-from-perfect characteristics of current quantum hardware. Consequently, the quest to uncover real-world quantum applications is not only a technical endeavor, but it may need a profound rethinking of how to tackle the problem itself.

In parallel, Artificial Intelligence (AI) has brought about a second transformative wave in the realm of computation. AI offers practical solutions to a wide range of complex problems, ranging from image recognition and natural language processing to autonomous driving and personalized medicine. However, the practical success of AI often comes without provable theoretical guarantees, making it challenging for the more traditional ways of studying computation through complexity theory and algorithms with worst-case performance guarantees. Machine learning systems derive their power from the ability to learn from data rather than following explicit, rule-based algorithms, and through techniques such as neural networks, reinforcement learning, and generative models, they continuously improve by adapting to new information, identifying patterns, and making decisions based on experience. This self-learning approach allows AI to provide practical solutions to problems that are difficult or even impossible to solve with traditional methods and provable guarantees. However, this also introduces a level of unpredictability and opacity that distinguishes it from classical algorithmic solutions.

\newpage
Among the many approaches in AI, one particularly relevant to our work is Reinforcement Learning (RL), where agents learn to make decisions by interacting with an environment to maximize their reward. DeepMind has utilized RL to achieve remarkable successes, starting with AlphaGo \cite{AlphaGo}, where the agent learned to play the game of Go at a superhuman level, and its successors, including AlphaZero and MuZero \cite{AlphaZero,muZero}, who mastered a large variety of ever more complicated games. In the realm of scientific research, DeepMind's AlphaFold \cite{AlphaFold} has made groundbreaking contributions to chemistry and biology by tackling the protein folding problem, for which it was recently awarded the 2024 Nobel Prize in Chemistry, while in mathematics, AI has already helped humans to prove new theorems \cite{DMMath}. Last, OpenAI's ChatGPT and other LLM systems benefit from RL through a process called Reinforcement Learning from Human Feedback (RLHF). All these examples push the boundaries of what it means to compute, favoring a more open-ended computational process of agents that perceive their environment and learn how to take actions that optimize their rewards.   

These developments suggest an opportunity at the intersection of quantum computing and AI: Quantum Intelligence. We use this term to denote agents that are now enhanced with quantum technologies as they perceive their environment and optimize their actions. Such agents could use a combination of quantum technologies for different tasks: quantum sensors to perceive quantum effects in their environment, quantum communication to interact in a distributed environment, parameterized quantum circuits as policy neural networks, quantum computers to perform simulations of quantum systems, or quantum-inspired ideas performed on classical computers. 

Quantum Intelligence extends previous work on quantum machine learning, a prolific area of quantum computing and still one of the promising areas for quantum applications \cite{kerenidis2016quantum, kerenidis2019qmeans, Landman2022quantummethods, kazdaghli2023improved, cherrat2022quantum, NearestCentroid2021, KP2022SubspaceStates, QNN2020, QCNN2019, kerenidis2018sfa,Cherrat2023,Thakkar2024,Jain2024}, and quantum reinforcement learning \cite{Chen2020, Skolik2022, Wang21, kerenidis2023}, by emphasizing autonomous, interactive, and adaptive behavior in quantum-native settings. Our focus is on creating a general and flexible framework in which quantum intelligent agents can be trained to discover, through interaction alone, strategies and algorithms of comparable quality to those designed by human experts. We formalize this framework in Section~\ref{sec2}.

Our contribution is a unified agent–environment formalism that encompasses both single- and multi-agent quantum tasks with shared registers and multi-round interactions. Training proceeds as direct policy search from episodic, measurement-defined rewards—without target circuits or gradients of known formulas—so agents learn solely from outcomes available on hardware. Policies are parameterized circuits constrained to nearest-neighbor connectivity and shallow depth, yielding human-readable structures that generalize beyond the instances seen in training. Crucially, the same learning loop runs unchanged on simulators for tiny instances and transitions seamlessly to quantum devices once minimum instance sizes exceed classical simulability, providing a clear simulator to device path.

\subsection{Our Results}

We present a general framework for quantum intelligent agents and demonstrate their ability to autonomously rediscover a range of foundational quantum algorithms and interactive quantum protocols. The framework models interactions between agents and quantum environments—or between multiple agents—through shared quantum registers, with policies implemented as parameterized quantum circuits. It specifies the key elements of these interactions, including private and shared registers, unitary operations representing agent policies, and measurement processes determining rewards. The design accommodates both single-agent and multi-agent settings, and is compatible with realistic hardware constraints such as nearest-neighbor connectivity.

To instantiate the framework, we define a versatile family of parameterized quantum circuits that can serve as expressive and trainable policy networks. Although we use a specific circuit family in our experiments, the framework is agnostic to the choice of architecture and can incorporate other quantum or hybrid models depending on the task.

We validate this approach through four representative examples:
\begin{enumerate}
	\item Quantum Fourier Transform (QFT) — learning efficient, logarithmic-depth nearest-neighbor circuits that implement (up to local phases and within numerical tolerance) the QFT.
	\item Grover’s Search Algorithm — rediscovering the optimal query algorithms for unstructured search, that start by the uniform superposition and apply a diffusion operator.
	\item Strong Quantum Coin Flipping — autonomously identifying cheating strategies matching the known optimal bias of protocols.
	\item CHSH and Conflicting-Interest Games — learning strategies achieving the optimal Tsirelson's bound, in both cooperative and competitive settings.
\end{enumerate}

In each case, the agents learn optimal strategies directly from interaction, without being given the known optimal circuits or measurements, and in a form that can be generalized to the standard algorithm. 

Our experimental validation uses few-qubit agents because the mathematical tasks we study admit meaningful instances at very small input sizes, and the learned algorithms then scale straightforwardly. The framework itself is scale-agnostic and explicitly aligned with hardware realities (e.g., nearest-neighbor connectivity and limited depth). For more complex tasks—such as portfolio optimization, where even the smallest practically relevant instance may require hundreds of qubits to encode decisions and constraints, or the search for hardware-tailored error-correction strategies on devices with hundreds to thousands of qubits—reward evaluation quickly exceeds classical simulability. In these regimes, quantum execution becomes necessary, and our learning loop transitions seamlessly from simulator to quantum hardware without modification. In short, small-scale demonstrations on clean mathematical problems establish correctness and interpretability, while the methodology is designed for—and becomes essential in—larger-scale settings where quantum resources are indispensable.

These results serve as a proof of principle that quantum intelligent agents can act as effective tools for algorithmic discovery, complementing rather than replacing human expertise, and opening new directions for the automated development of quantum algorithms and protocols. 

\subsection{Related work}

A growing line of “algorithm discovery’’ in quantum computing uses learning-driven methods to recover short-depth circuits or even analytic forms. Cincio \emph{et al.} \cite{Cincio_2018} learned compact circuits for estimating state overlap and then extrapolated the learned structures into provable analytic algorithms, illustrating how machine-learned circuits can yield human-verifiable insights. 
Morales \emph{et al.} \cite{PhysRevA.98.062333} variationally learned Grover’s search, recovering optimal performance (and small-size improvements) while confirming that variational training can rediscover canonical oracle algorithms. 
Variational Quantum Cloning (VQC) similarly frames cloning as a learnable objective, combining operationally meaningful costs, gradient-based training, and circuit-structure learning targeted to NISQ constraints \cite{PhysRevA.105.042604}. 
Beyond task-specific exemplars, ADAPT-VQE adaptively grows ansätze from operator pools to minimize energy, embodying a powerful but domain-specific (Hamiltonian-targeted) form of circuit structure discovery \cite{GEB2019}. 
In parallel, quantum architecture search (QAS) automates PQC design with differentiable, evolutionary, or noise-aware co-search—e.g., DQAS and QuantumNAS—and recent surveys consolidate methodology and scope \cite{Zhang_2022,Wang_2022,Martyniuk_2024}. 

Our contribution differs in framing algorithmic discovery as episodic interaction within a unified agent–environment formalism (single- and multi-agent), covering both algorithmic primitives (e.g., QFT, Grover) and interactive protocols/games (e.g., coin-flipping, CHSH), with policies constrained to nearest-neighbor hardware and rewards defined beyond energy minimization—thereby complementing prior approaches while emphasizing generality and interactivity. 

\begin{figure}[t!]
	\centering
	\includegraphics[width=0.95\textwidth]{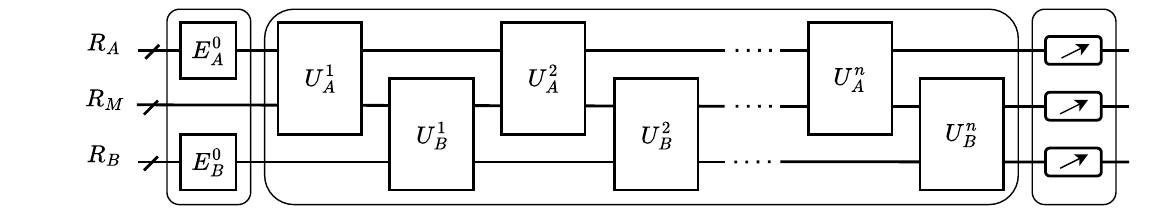}
	\caption{  General framework for quantum environments involving two agents, $A$ and $B$. 
		The system consists of three registers: $R_A$ (private to $A$), $R_B$ (private to $B$), and $R_M$ (message register shared between them). 
		The environment prepares the initial state via $E_A^0$ and $E_B^0$. 
		In round $t$ of an interaction, $A$ applies $U_A^t$ to $R_A \otimes R_M$, followed by $B$ applying $U_B^t$ to $R_M \otimes R_B$. 
		After all rounds of the interaction, the registers are measured. 
		An episode consists of multiple interactions, potentially with different inputs, and the final reward is computed from all measurement outcomes in the episode.
	}
	\label{fig1}
\end{figure}

\section{Quantum Intelligent Agents}\label{sec2}

\subsection{Framework for Agent Interactions}\label{sec:framework}

We introduce a general quantum learning framework that models interactions between quantum agents and their environment, formalized within the agent-environment paradigm of reinforcement learning and the mathematical structure of quantum communication protocols. In this framework, agents exchange quantum information via shared registers and act on private registers using learnable unitary policies.

An \emph{interaction} consists of a fixed number $T$ of \emph{rounds} during which agents apply unitary operations to their private registers and to a shared \emph{message register}. The inputs to all registers remain fixed throughout a single interaction. At the conclusion of the interaction, all registers are measured in the computational basis, producing classical bit strings.

An \emph{episode} is a sequence of $K$ interactions. Between interactions, the environment may provide different inputs sampled from a specified distribution. A scalar reward is computed at the end of the episode from a problem-specific utility function applied to all measurement outcomes within that episode. Formally, an interaction operates on fixed inputs across multiple rounds before measurement, whereas an episode comprises multiple interactions with varying inputs before reward computation.

We consider two primary instantiations of this framework (Fig.~\ref{fig1}):
\begin{enumerate}
	\item \textbf{Two-agent setting} — agents $A$ and $B$ share a message register $R_M$ while maintaining private registers $R_A$ and $R_B$. The environment supplies initial states and computes rewards.
	\item \textbf{Single agent with quantum environment} — agent $A$ maintains a private register $R_A$ and shares a message register $R_M$ with the environment $E$, which maintains a private register $R_B$, provides inputs, and computes rewards through direct participation via the shared register $R_M$.
\end{enumerate}

For two agents $A$ and $B$, the global system contains
\[
N = N_A + N_M + N_B
\]
qubits, partitioned into:
\begin{itemize}
	\item $R_A$: private register of $A$ ($N_A$ qubits)
	\item $R_B$: private register of $B$ or the environment ($N_B$ qubits)
	\item $R_M$: message register shared between them ($N_M$ qubits)
\end{itemize}

At the start of an interaction, the environment prepares the initial state of all registers via unitaries $E_A^0$ and $E_B^0$, encoding problem inputs on $R_A$ and $R_B$. For each round $t \in \{1, \ldots, T\}$ within that interaction:
\begin{enumerate}
	\item Agent $A$ applies a unitary $U_A^t$ to $R_A \otimes R_M$
	\item Agent $B$ applies a unitary $U_B^t$ to $R_M \otimes R_B$
\end{enumerate}

Each agent maintains a policy $\pi_A$ (or $\pi_B$) that selects the unitary $U_A^t$ (or $U_B^t$) at round $t$ from an available policy set, which may be fixed or parameterized for learning. After round $T$, the registers $R_A$, $R_M$, and $R_B$ are measured in the computational basis, yielding classical outcomes associated with that interaction.

An episode consists of $K$ such interactions, potentially with different inputs chosen by the environment at the start of each interaction. The environment computes the final reward by applying a utility function to all measurement outcomes from the episode. This reward is then used to update the agents' policy parameters.

We train policies via episodic direct policy search: in each episode, the environment samples inputs, the agent applies its policy unitaries across all rounds and interactions, and a single scalar reward is obtained from quantum measurements (e.g., state fidelity, success probability, or game payoff). We do not provide supervised labels or target parameters; learning proceeds solely from trial-and-error optimization of measured returns. On quantum hardware, rewards are estimated from finite-shot sampling and can incorporate readout error mitigation.

In our implementation, policies are parameterized quantum circuits whose parameters are optimized via gradient-based methods to maximize expected rewards over episodes. This design is compatible with near-term quantum devices. More generally, the framework accommodates arbitrary unitary families, classical control logic, or hybrid quantum-classical policies.

This framework naturally instantiates diverse interaction scenarios:

\begin{itemize}
	\item A single-round interaction between one quantum agent and a fixed-policy environment (agent seeks a quantum algorithm for a computational problem; environment provides inputs and computes rewards).
	\item Multi-round interactions between a quantum agent and a fixed-policy environment (agent seeks an optimal strategy in a sequential game, e.g., coin-flipping protocols).
	\item Multi-round interactions between multiple learning agents, where all agents simultaneously adapt their policies (agents jointly discover optimal strategies for nonlocal games).
\end{itemize}

By appropriately specifying the reward function and the policy set, this framework enables systematic exploration of quantum algorithms, cryptographic protocols, and quantum games. Detailed instantiations are provided in Section~\ref{sec:examples}.

\begin{figure}[t!]
	\centering
	\includegraphics[width=0.9\textwidth]{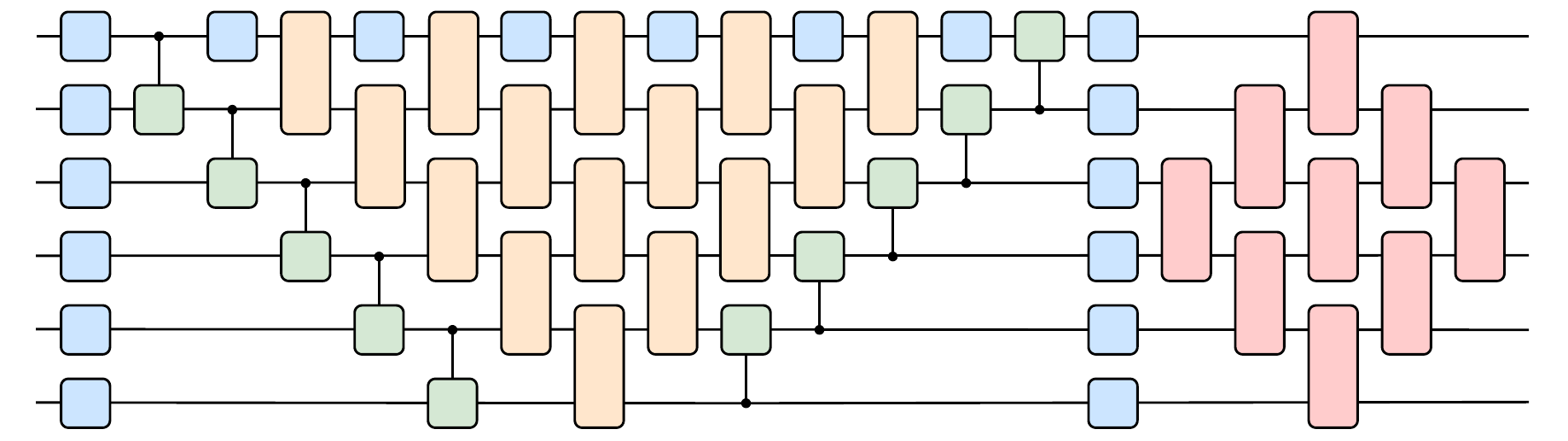}
	\caption{Example parameterized quantum circuit serving as an agent's policy network. Blue: $U(\theta, \phi)$ single-qubit gates; green: $CRY(\theta)$ controlled rotations; orange: $M(\theta, \phi_1, \phi_2)$ matchgates; red: $M(\theta, \pi, 0)$ matchgates used for SWAP operations. This structure balances expressivity, trainability, and nearest-neighbor hardware constraints.}
	\label{fig2}
\end{figure}

\subsection{Parameterized Quantum Circuits as Agent Policies}

In our framework, an agent's policy is a set of unitary operations it can apply during each round of an interaction (see Section~\ref{sec:framework}). At every round, the agent selects a unitary from this set to act on its accessible registers, which may include both private qubits and shared message qubits. Each unitary corresponds to a specific policy action available to the agent at that round.

To enable learning, we represent these unitaries as \emph{parameterized quantum circuits} (PQCs) with adjustable parameters that are optimized during training. PQCs provide a flexible and expressive way to model agent policies while accommodating hardware constraints such as nearest-neighbor connectivity.

\paragraph{Input Encoding.}  
The framework supports both quantum and classical inputs. Classical binary data $c \in \{0,1\}$ is encoded as $\ket{c}$ in a single qubit, while real-valued vectors $x \in \mathbb{R}^n$ can be amplitude-encoded as
\[
\ket{x} = \frac{1}{\|x\|} \sum_{i=1}^n x_i \ket{i}
\]
over $\lceil \log_2 n \rceil$ qubits. Alternative encodings (e.g., basis encoding, angle encoding) may be employed as appropriate for specific problem structures. Inputs are either deterministic or stochastically sampled according to the problem specification.

\paragraph{Parameterized Gates.}  
The PQCs are constructed from the following learnable gates, each designed to balance expressivity and hardware feasibility:

\begin{enumerate}
	\item \textbf{Single-qubit rotation and phase shift gate}:
	
	\begin{equation} \label{U}
		\text{U}(\theta,\phi) = \begin{pmatrix}
		\cos \frac{\theta}{2} & - e^{i \phi} \sin \frac{\theta}{2} \\
		\sin \frac{\theta}{2} & e^{i \phi} \cos \frac{\theta}{2}
		\end{pmatrix}
	\end{equation}
	
	This gate is mathematically equivalent to the composition of a Y-axis rotation and a phase shift:
	\[
	\text{U}(\theta,\phi) = \text{RY}(\theta) \cdot \text{PhaseShift}(\phi)
	\]
	
	Special cases include:
	$\text{U}(\theta,0) = \text{RY}(\theta)$,  
	$\text{U}(0,\phi) = \text{PhaseShift}(\phi)$,  
	$\text{U}\left(\frac{\pi}{2},\pi\right) = \text{H}$ (Hadamard gate),  
	$\text{U}(\pi,\pi) = \text{X}$ (Pauli-X gate).
	
    \item \textbf{Two-qubit matchgate}:
	      
	      \begin{equation} \label{M}
	      	\text{M}(\theta, \phi_1, \phi_2) = \left( \begin{array}{cccc}
	      	1 & 0 & 0 & 0 \\
	      	0 & \cos \frac{\theta}{2} & - e^{i \phi_1} \sin \frac{\theta}{2} & 0 \\
	      	0 & \sin \frac{\theta}{2} & e^{i \phi_1} \cos \frac{\theta}{2} & 0 \\
	      	0 & 0 & 0 & e^{i\phi_2} \end{array} \right)
	      \end{equation}
	      
	      This gate can be decomposed into a two-qubit rotation and phase shifts. Matchgates can implement fermionic operations and, with specific parameters, SWAP gates ($M(\pi,\pi,0) = \text{SWAP}$).
          
	\item \textbf{Controlled-$R_Y$ rotation}:
	
	\begin{equation} \label{CU}
		\text{CRY}(\theta) = \begin{pmatrix}
		1 & 0 & 0 & 0 \\
		0 & 1 & 0 & 0 \\
		0 & 0 & \cos \frac{\theta}{2} & - \sin \frac{\theta}{2} \\
		0 & 0 & \sin \frac{\theta}{2} & \cos \frac{\theta}{2}
		\end{pmatrix}
	\end{equation}
	
	This gate provides conditional rotations necessary for generating entanglement between qubits.
\end{enumerate}

Each gate instance in the circuit has independent learnable parameters $\theta$ and $\phi$, so the total number of parameters scales with the circuit size.

\paragraph{Parameterized Quantum Circuit Architecture.}  
The circuit architecture consists of six sequential layers, constrained to nearest-neighbor connectivity:

\begin{enumerate}
	\item \textbf{RYPhaseShift layer:} $\text{U}(\theta,\phi)$ applied to each qubit.
	\item \textbf{CRY downward ladder:} $\text{CRY}(\theta)$ gates between consecutive qubits $(i,i+1)$.
	\item \textbf{Enhanced matchgate pyramid:} $\text{M}(\theta,\phi_1,\phi_2)$ gates in a pyramidal pattern, with $\text{U}(\theta,\phi)$ gates interleaved on the first qubit.
	\item \textbf{CRY upward ladder:} $\text{CRY}(\theta)$ gates applied in reverse order $(i+1,i)$.
	\item \textbf{RYPhaseShift adjoint layer:} $\text{U}(\theta,\phi)^\dagger$ applied to all qubits.
	\item \textbf{SWAP layer (optional):} $\text{M}(\theta,\pi,0)$ gates arranged to swap private and shared registers when $\theta = \pi$, used in multi-round protocols.
\end{enumerate}

For $n$ qubits, this architecture contains $O(n^2)$ two-qubit gates and $O(n)$ single-qubit gates per layer, yielding $O(n^2)$ total learnable parameters.

\paragraph{Remarks.}  
\begin{itemize}
	\item \textbf{Parameter scaling:} The $O(n^2)$ parameterization provides sufficient expressivity under nearest-neighbor connectivity constraints.
	
	\item \textbf{Expressivity and trainability:} This circuit family provides a balance between expressivity and trainability. Alternative architectures may be substituted depending on task requirements.
	
	\item \textbf{Depth extension:} The architecture is analogous to a single-layer quantum neural network; additional layers may be stacked to increase depth.
	
	\item \textbf{Fermionic capability:} Specific parameter settings yield fermionic circuits; when combined with $\text{U}$ and $\text{CRY}$ gates in the beginning of the circuit, these can be classically hard to simulate~\cite{Fermion}.
	
	\item \textbf{Interactive communication:} The trainable SWAP operation enables register exchange between private and message qubits in multi-round protocols.
	
	\item \textbf{Simplification:} Circuits can be pruned by removing layers or fixing parameters, enabling post-training simplification and analysis of learned structures.
\end{itemize}

In summary, the parameterized quantum circuit defines the learnable unitary $U_A^t$ (or $U_B^t$) used by the agents at each round of an interaction, with parameters optimized to maximize expected episodic reward.

\section{Examples}\label{sec:examples}

We instantiate our framework through four canonical quantum tasks spanning algorithmic primitives, cryptographic protocols, and nonlocal games. In each case, agents learn optimal or near-optimal strategies solely from episodic rewards, without access to known solutions or structural guidance. The learned circuit structures exhibit interpretable patterns that enable systematic generalization to arbitrary problem sizes.

All experiments used quantum simulators with training durations under 10 minutes on standard hardware. Implementation employed PyTorch~\cite{paszke_pytorch:_2019} for optimization, PennyLane~\cite{bergholm_pennylane:_2022} for differentiable circuit execution, and Qiskit~\cite{aleksandrowicz_qiskit:_2019} for circuit construction and visualization. Each result was validated over multiple random seeds to ensure robustness. Complete experimental specifications enable reproducibility.

\begin{figure}[t]
	\centering
    \includegraphics[width=0.55\textwidth]{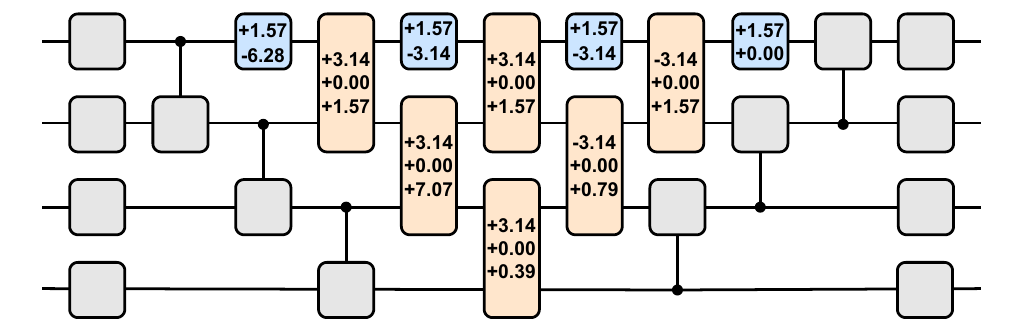}
	\caption{Learned QFT circuit for $n=4$. Only the enhanced matchgate pyramid layer contains non-zero parameters.}
	\label{QFT1}
\end{figure}

\begin{figure}[ht]
	\centering
	\includegraphics[width=0.7\textwidth]{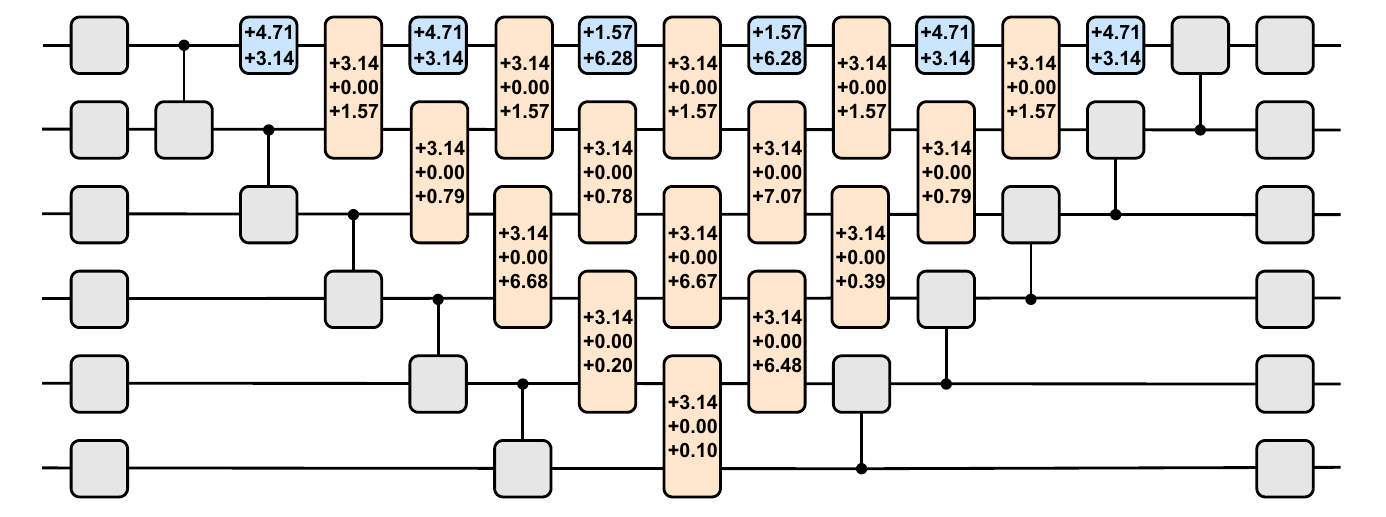}
	\caption{Learned QFT circuit for $n=6$.}
	\label{QFT2}
\end{figure}

\subsection{The Quantum Fourier Transform}

\paragraph{Problem Description and Objective.} 
The Quantum Fourier Transform (QFT) is the quantum analogue of the Discrete Fourier Transform, acting on a $2^n$-dimensional vector encoded in $n$ qubits. It can be implemented with a polynomial-size quantum circuit, providing exponential speedup over the classical DFT, which requires $\tilde{O}(2^n)$ time. QFT is a key subroutine in quantum factoring and phase estimation algorithms. Our objective is to train an agent to discover a polynomial-size QFT circuit respecting nearest-neighbor connectivity constraints.

\paragraph{Environment Definition.}
We use a simplified interaction architecture where both agent and environment act only on a shared register $R_M$ with $n$ qubits. The agent applies a parameterized quantum circuit constrained to nearest-neighbor connectivity. The environment prepares computational basis states $\ket{x}$ with $x \in \{0,1\}^n$. Training on basis states suffices, since correctness extends by linearity to arbitrary superpositions. The target output for input $\ket{x}$ is
\[
\ket{\psi_{\text{out}}} = \text{QFT}_{2^n} \ket{x} = \frac{1}{\sqrt{2^n}} \sum_{k=0}^{2^n-1} e^{2\pi i x k / 2^n} \ket{k}.
\]

\paragraph{Training Protocol.} 
Each episode consists of $2^n$ interactions, one per basis state $x \in \{0,1\}^n$. For each interaction: (i) the environment prepares $\ket{x}$, (ii) the agent applies its policy circuit, (iii) the environment computes fidelity to the ideal QFT output. The reward $R$ is the average probability of measuring the all-zero string after applying inverse QFT and inverse state preparation to the agent's output, effectively measuring $\ell_2$ distance to the target state. Fidelity computation can be performed via state tomography on quantum hardware.

\paragraph{Results.} 
Training converged after 300 epochs for both $n=4$ and $n=6$ qubits. For $n=4$, the agent achieved fidelity $\mathbf{0.999999}$ (Fig.~\ref{QFT1}). For $n=6$, fidelity reached $\mathbf{0.999999}$ (Fig.~\ref{QFT2}). Circuit simplification revealed that only the Enhanced Matchgate Pyramid layer required non-zero parameters; all other layers could be zeroed without performance degradation.

\begin{figure}[t]
	\centering
    \includegraphics[width=0.9\textwidth]{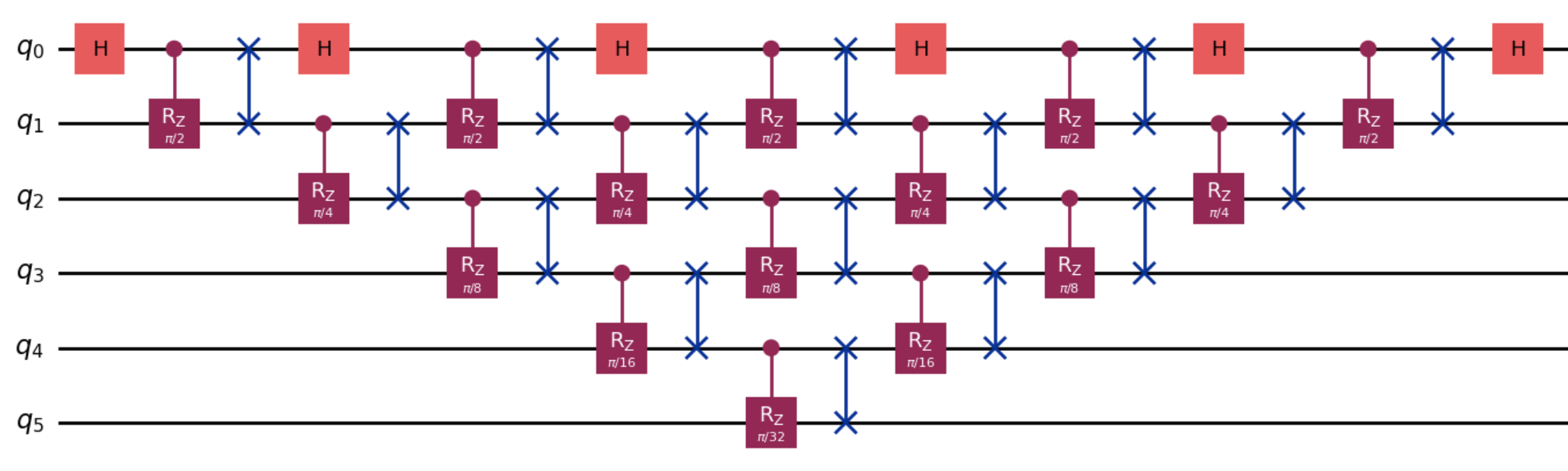}
	\caption{The general QFT algorithm optimized for nearest-neighbor connectivity. The circuit uses $O(n^2)$ gates and depth $O(n)$, implementing the pattern learned by the quantum agents.}
	\label{QFT}
\end{figure}

\paragraph{Analysis.} 
Examination of the learned circuits reveals a consistent structure: single-qubit gates on the first qubit of the pyramid implement Hadamard operations (up to global phases), while matchgates perform controlled-$R_z$ rotations with cascading phase shifts $\pi/2^k$ for $k = 1, 2, \ldots$, followed by SWAP operations. This pattern generalizes to arbitrary $n$ as follows: at depth level $d \in \{0, 1, \ldots, n-2\}$, the circuit applies a Hadamard gate to qubit $d$, then performs controlled-$Z$ rotations between qubit $d$ and qubits $d+1, d+2, \ldots, n-1$ with phases $\pi/2^k$ for $k = 1, \ldots, n-d-1$, followed by SWAP operations propagating the qubit to position $n-1$. This nearest-neighbor implementation achieves optimal depth $O(n)$ and size $O(n^2)$, using $\frac{1}{2} n (n - 1)$ controlled rotations, $\frac{1}{2} n (n - 1)$ SWAP gates, and $n$ single-qubit gates (Fig.~\ref{QFT}).

\newpage

\subsection{Quantum Coin Flipping}

\paragraph{Problem Description and Objective.} 
Quantum coin flipping is a cryptographic primitive enabling two distrustful parties to generate a random bit over distance. A strong coin flipping protocol with bias $\varepsilon$ (denoted $\text{SCF}(\varepsilon)$) satisfies: (i) if both parties are honest, the outcome $c$ satisfies $\Pr(c = 0) = \Pr(c = 1) = 1/2$; (ii) if Alice cheats and Bob is honest, $P^*_A = \max\{\Pr(c = 0), \Pr(c = 1)\} \le 1/2 + \varepsilon$; (iii) if Bob cheats and Alice is honest, $P^*_B \le 1/2 + \varepsilon$. We consider a protocol with bias $\varepsilon = 1/4$ based on~\cite{kerenidis2004,ambainis2001}, for which optimal cheating strategies achieve $P^* = 3/4$.

The protocol proceeds as follows: (1) Alice randomly selects $a \in \{0,1\}$, prepares the qutrit pair state $\ket{\psi_a} = \frac{1}{\sqrt{2}}(\ket{aa} + \ket{22})$, and sends Bob the second qutrit. (2) Bob randomly selects $b \in \{0,1\}$ and sends it to Alice. (3) Alice reveals $a$ and sends her remaining qutrit. Bob verifies the state via projective measurement in the basis $\{\ket{\psi_a}, \ket{\psi_a}^\perp\}$. If verification passes, the outcome is $c = a \oplus b$; otherwise, Bob aborts.

Optimal cheating strategies are known analytically: Bob measures his received qutrit and sets $b$ to force outcome $c$ if he measures $a$, or chooses $b$ randomly if he measures $2$, achieving $P^*_B = \frac{1}{2} + \frac{\|\rho_0 - \rho_1\|}{4} = 3/4$, where $\rho_0, \rho_1$ are reduced density matrices of Bob's qutrit when $a=0$ or $a=1$. Alice's optimal strategy prepares the state $\ket{\psi^d} = \frac{1}{\sqrt{6}}(\ket{00} + \ket{11} + 2\ket{22})$ instead of $\ket{\psi_a}$, again achieving $P^*_A = 3/4$. Our objective is to determine whether agents can autonomously discover strategies achieving this optimal bias without prior knowledge of these solutions.

\begin{figure}[t]
	\centering
	\includegraphics[width=0.45\textwidth]{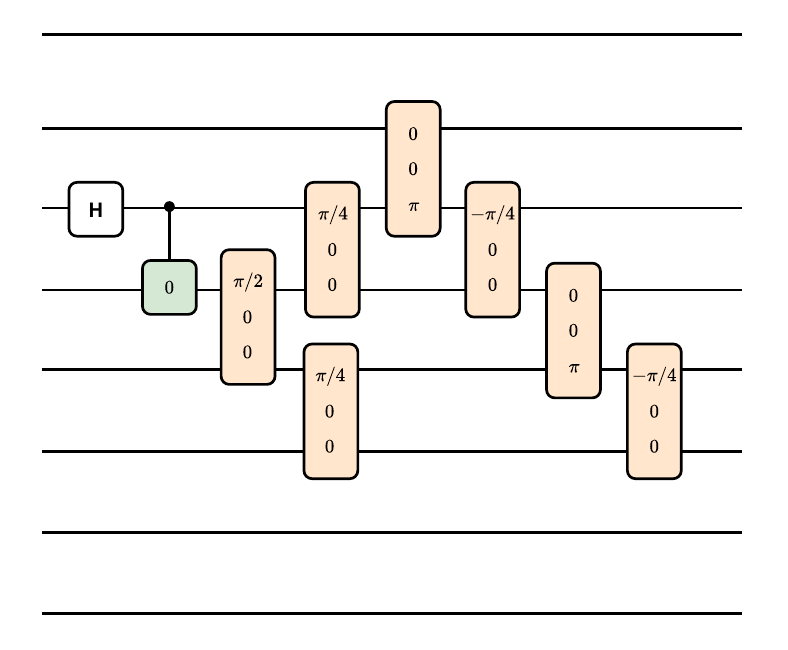}
	\caption{Circuit implementing the honest protocol's main unitary operation.}
	\label{fig4}
\end{figure}

\paragraph{Implementation.} 
We encode qutrits using two qubits: $|0\rangle \to \ket{10}$, $|1\rangle \to \ket{01}$, $|2\rangle \to \ket{00}$. The honest protocol can be expressed within our framework using a 12-qubit system with registers $R_A$, $R_M$, $R_B$ (4 qubits each). Figure~\ref{fig4} shows a specific parameterization implementing the honest protocol's state preparation. Agent $A$ creates the state $\ket{a}(\frac{1}{\sqrt{2}}\ket{0000} + \frac{1}{\sqrt{2}}\ket{aa})$ on its register and message qubits, which is then shared with agent $B$ via register swapping. Agent $B$ performs the adjoint operation and verifies the state. Measurement of the first two qubits in each agent's register yields their respective coin values, with outcome $c = a \oplus b$.

\paragraph{Environment Definition.}
The interaction uses the two-round architecture from Fig.~\ref{fig1}, where Alice applies unitaries $\{U_A^1, U_A^2\}$ and Bob applies $\{U_B^1, U_B^2\}$. No external inputs are provided; honest agents internally initialize with random bits $a, b \in \{0,1\}$. The outcome is $c = a \oplus b$, which agents aim to bias when cheating. To train cheating strategies, we fix one agent's policy to the honest protocol and optimize the other agent's policy. Agents may implement randomized strategies by conditioning on an internal coin flip.

\paragraph{Training Protocol.}
Each episode contains two interactions corresponding to the honest agent's coin values $a = 0$ and $a = 1$ (or $b = 0$ and $b = 1$ when training Alice). The reward is the cheating agent's probability of achieving their desired outcome, averaged over both interactions. The target cheating probability is $3/4$.

\paragraph{Results.} 
Training used a 12-qubit system (4 qubits per register) and converged after 300 epochs. The trained cheating agent for Alice achieved probability $\mathbf{P^*_A = 0.749992}$ (optimal: $0.750000$). The trained cheating agent for Bob achieved $\mathbf{P^*_B = 0.749985}$ (optimal: $0.750000$). The circuits appear in the Appendix since they are pretty large, given they have 12 qubits. 

\newpage

\paragraph{Analysis.}
Agents successfully learned optimal cheating strategies without knowledge of the analytically derived solutions. The parameterized circuits proved sufficiently expressive to capture complex quantum cryptographic protocols, including adversarial deviations from honest behavior. This demonstrates the framework's capability to discover attack strategies in multi-round interactive protocols where analytical characterization may be intractable.

\subsection{The CHSH and Other Nonlocal Games}

\paragraph{Problem Description and Objective.} 
We consider the CHSH game and a conflicting-interest variant combining CHSH with the Battle of the Sexes~\cite{pappa2015}. Following~\cite{pappa2015}, we define a two-party Bayesian game with: (i) players Alice ($A$) and Bob ($B$); (ii) type sets $\mathcal{X}_A, \mathcal{X}_B$ with joint distribution $P: \mathcal{X}_A \times \mathcal{X}_B \to [0,1]$; (iii) action sets $\mathcal{Y}_A, \mathcal{Y}_B$; (iv) utility functions $u_i: \mathcal{X}_A \times \mathcal{X}_B \times \mathcal{Y}_A \times \mathcal{Y}_B \to \mathbb{R}$ for $i \in \{A, B\}$. Each player receives type $x_i$ according to $P$, then independently selects action $y_i$ to maximize average payoff
\begin{equation}\label{avgpayoff}
F_i = \sum_{x \in \mathcal{X}} \sum_{y \in \mathcal{Y}} P(x) \Pr(y|x) u_i(x, y),
\end{equation}
where $\Pr(y|x)$ depends on the players' strategies.

\begin{table}[h!]
\centering
\begin{subtable}[t]{0.45\textwidth}
\centering
\begin{tabular}{|c|c|c|}
\hline
\backslashbox{$y_A$}{$y_B$} & \makebox[5em]{0} & \makebox[5em]{1} \\ \hline
0 & (1, 1/2) & (0, 0) \\ \hline
1 & (0, 0) & (1/2, 1) \\ \hline
\end{tabular}
\caption{$x_A \land x_B = 0$}
\label{table1}
\end{subtable}%
\hspace{1cm}
\begin{subtable}[t]{0.45\textwidth}
\centering
\begin{tabular}{|c|c|c|}
\hline
\backslashbox{$y_A$}{$y_B$} & \makebox[5em]{0} & \makebox[5em]{1} \\ \hline
0 & (0, 0) & (3/4, 3/4) \\ \hline
1 & (3/4, 3/4) & (0, 0) \\ \hline
\end{tabular}
\caption{$x_A \land x_B = 1$}
\label{table2}
\end{subtable}
\caption{Payoff matrices for the conflicting-interest game. Utilities depend on the logical AND of player types and differ between players for $x_A \land x_B = 0$, creating conflicting interests.}
\label{tab:payoff}
\end{table}

The conflicting-interest game utilities (Table~\ref{tab:payoff}) exhibit coordination requirements similar to CHSH except when $x_A \land x_B = 1$, where anti-coordination is required. Unlike standard CHSH, players have divergent preferences when $x_A \land x_B = 0$: Alice prefers $(0,0)$ while Bob prefers $(1,1)$. The optimal quantum strategy shares a maximally entangled state $\ket{\Phi} = \frac{1}{\sqrt{2}}(\ket{00} + \ket{11})$ and employs projective measurements
\begin{align*}
\mathcal{A}_0^a &= \ketbra{\phi_a(0)}{\phi_a(0)}, \quad \mathcal{A}_1^a = \ketbra{\phi_a(\pi/4)}{\phi_a(\pi/4)}, \\
\mathcal{B}_0^b &= \ketbra{\phi_b(\pi/8)}{\phi_b(\pi/8)}, \quad \mathcal{B}_1^b = \ketbra{\phi_b(-\pi/8)}{\phi_b(-\pi/8)},
\end{align*}
where $\phi_0(\theta) = \cos\theta\ket{0} + \sin\theta\ket{1}$ and $\phi_1(\theta) = -\sin\theta\ket{0} + \cos\theta\ket{1}$. This yields $\Pr(y_A, y_B | x_A, x_B) = \frac{1}{2}\cos^2(\pi/8)$ and average payoff $F_i = \frac{3}{4} \cdot 0.85 \approx 0.640165$ for both players, which is the optimal quantum equilibrium~\cite{pappa2015}.

\begin{figure}[t]
	\centering
	\includegraphics[width=0.9\textwidth]{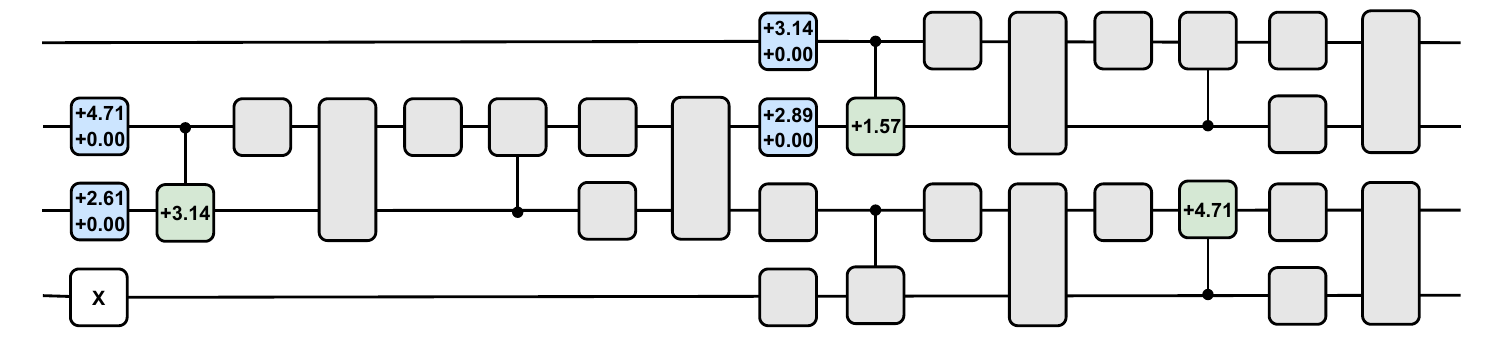}
	\caption{Trained quantum agents for the CHSH and conflicting-interest games. Optimal strategies are achieved using only RYPhaseShift layers and CRY ladders. Here the environment has provided inputs $x_A=0, x_B=1$}
	\label{CHSH}
\end{figure}

\begin{figure}[h]
	\centering
	\includegraphics[width=0.35\textwidth]{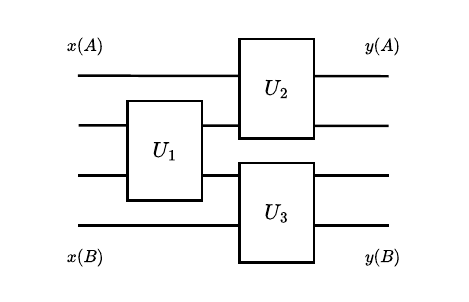}
	\caption{Interaction architecture for nonlocal games.}
	\label{fig5}
\end{figure}

\paragraph{Environment Definition.}
The interaction architecture (Fig.~\ref{fig5}) ensures input-independent entanglement generation by restricting agent $A$'s first unitary to not access the qubit containing input $x_A$. Agents receive uniformly random inputs $x_A, x_B \in \{0,1\}$ encoded as basis states $\ket{x_A}$, $\ket{x_B}$ in registers $R_A$, $R_B$. Agent $A$ applies a parameterized circuit before input distribution, then both agents apply input-dependent local circuits before measuring in the computational basis to produce outputs $y_A, y_B \in \{0,1\}$.

\paragraph{Training Protocol.}
Each episode contains four interactions, one per input pair $(x_A, x_B) \in \{0,1\}^2$. The reward is the average payoff $F_i$ over all four interactions, computed according to Eq.~\eqref{avgpayoff}.

\paragraph{Results.}
Both agents trained simultaneously in a 4-qubit system for 300 epochs. For standard CHSH, agents achieved average payoff $\mathbf{F = 0.853553}$ (optimal: $\cos^2(\pi/8) \approx 0.853553$). For the conflicting-interest game, agents achieved $\mathbf{F = 0.640154}$ (optimal: $\frac{3}{4}\cos^2(\pi/8) \approx 0.640165$). Circuit simplification revealed that only RYPhaseShift layers and CRY ladders were necessary; all other layers could be removed without performance loss (Fig.~\ref{CHSH}).

\paragraph{Analysis.}
Agents autonomously discovered strategies reproducing optimal quantum correlations for both games without prior knowledge of entangled states or measurement settings. The learned state preparation differs from the canonical maximally entangled state by local transformations. Specifically, agent $A$ prepares (independent of input):
\[
\cos\left(\pi/8\right) \frac{1}{\sqrt{2}}(\ket{00} - \ket{11}) - \sin\left(\pi/8\right) \frac{1}{\sqrt{2}}(\ket{01} - \ket{10}),
\]
a weighted superposition of Bell states. Each agent then applies input-dependent local rotations: agent $A$ applies $X$ followed by $\text{CRY}(\pi/4)$ controlled by its input qubit, while agent $B$ applies $\text{CRY}(\pi/4)$ controlled by its input qubit. This policy achieves exactly $\cos^2(\pi/8)$ in standard CHSH and $\frac{3}{4}\cos^2(\pi/8)$ in the conflicting-interest variant, matching known optimal values. These results demonstrate successful multi-agent coordination in both cooperative and adversarial settings.

\subsection{Grover's Search Algorithm}

\paragraph{Problem Description and Objective.}
Grover's algorithm provides quadratic speedup for unstructured search, identifying a marked item $\ket{w}$ among $N$ candidates in $O(\sqrt{N})$ oracle calls. The algorithm requires two components:
\begin{itemize}[noitemsep]
    \item Initialization to uniform superposition $\ket{s} = \frac{1}{\sqrt{N}}\sum_{x=0}^{N-1} \ket{x}$ via Hadamard gates.
    \item Amplitude amplification via the Grover diffusion operator $2\ket{s}\bra{s} - \mathcal{I}$ applied after each oracle query.
\end{itemize}

\paragraph{Environment Definition.}
Our objective is to train an agent to recover this procedure solely from reward feedback.
The environment implements the oracle $O: \ket{x} \mapsto (-1)^{\delta_{xw}} \ket{x}$, which marks the target element $\ket{w}$ by phase flip. The agent applies parameterized circuits before and after oracle calls to amplify the target state's amplitude. No inputs are provided to the agent; initialization and amplification must be learned.

\paragraph{Training Protocol.}
Each episode contains $N$ interactions, with the environment marking a different database element in each interaction. The reward is the probability of measuring the marked element after all circuit operations.

\begin{figure}[t]
	\centering
	\includegraphics[width=0.9\textwidth]{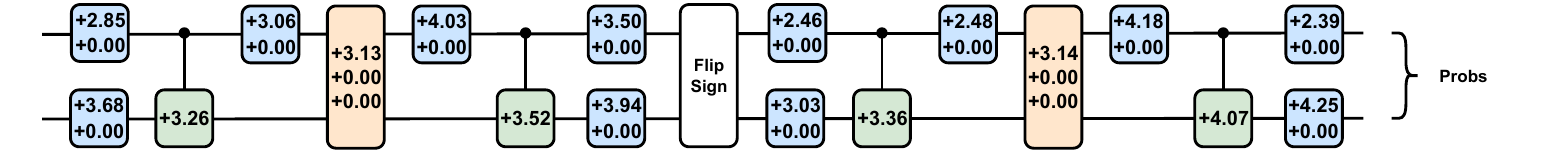}
	\caption{Trained circuit for Grover's search with one oracle call and database size $N=4$. The pre-oracle unitary is equivalent (up to phases) to tensor-product Hadamards.}
	\label{Grover1}
\end{figure}

\begin{figure}[t]
	\centering
\includegraphics[width=0.9\textwidth]{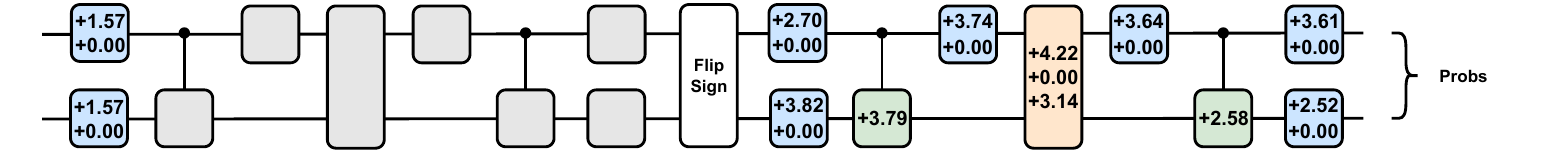}
	\caption{Trained circuit for Grover's search ($N=4$, one oracle call) with pre-oracle Hadamards fixed. The post-oracle unitary implements the diffusion operator using only nearest-neighbor two-qubit gates.}
	\label{Grover2}
\end{figure}

\paragraph{Results.}
We trained agents for database sizes $N \in \{4, 8\}$ with varying oracle calls. Training converged after 300 epochs for all cases.

\textit{Case $N=4$, one oracle call:} Initial training with one circuit layer plateaued; adding a second layer enabled the agent to achieve success probability $\mathbf{0.999999}$ (optimal: $1.000000$). Note also that we used a downward ladder with $CRY$ gates between $(i,i+1)$ qubits, but the same results could have retrieved if we reverse the control qubit of the gates.
Circuit analysis revealed that the pre-oracle operation is equivalent (up to phase shifts) to tensor-product Hadamards, creating uniform superposition. The post-oracle operation approximates the diffusion operator $2\ket{s}\bra{s} - \mathcal{I}$, sometimes preceded by an inconsequential phase flip on a single basis state (Fig.~\ref{Grover1}). 

To isolate the diffusion operator learning, we fixed the pre-oracle operation to explicit Hadamards and retrained only the post-oracle block, achieving success probability $\mathbf{0.999999}$. Unitary reconstruction confirmed that the learned operation decomposes as the product of the expected Grover diffusion operator and a phase flip on the third basis state. This phase flip does not affect search correctness and can be omitted, yielding the canonical Grover algorithm. The learned circuit respects nearest-neighbor connectivity and uses only two-qubit gates, avoiding multi-controlled gates typical in textbook implementations (Fig.~\ref{Grover2}).

\textit{Case $N=8$, one oracle call:} The agent achieved success probability $\mathbf{0.781249}$ (optimal: $0.781250$). Circuit analysis confirmed generalization of the $N=4$ solution: the pre-oracle block implements uniform superposition over 8 states, and the post-oracle block implements the corresponding diffusion operator.

\textit{Case $N=8$, two oracle calls:} Using transfer learning, we fixed the first two operations (uniform superposition and first diffusion) from the one-query case and trained the final post-oracle block. The agent achieved $\mathbf{0.944770}$ (optimal: $0.945313$). Composing the learned operation with the ideal diffusion operator yielded approximately the identity matrix (diagonal elements $>0.96$), indicating equivalence to a valid Grover iteration.

\paragraph{Analysis.}
The agent autonomously recovered both core components of Grover's search. In all tested cases, learned policies matched optimal success probabilities to numerical precision despite hardware-style constraints (nearest-neighbor connectivity, no multi-controlled gates). Circuit analysis confirmed functional equivalence to Grover's operators up to local phases, with demonstrated generalization to larger search spaces and multiple queries. The pattern generalizes naturally to arbitrary database sizes: extend the uniform superposition to $\lceil \log_2 N \rceil$ qubits, apply the corresponding diffusion operator, and repeat after each oracle query. The ability to rediscover a query-optimal algorithm solely from reward signals suggests potential extension to more complex oracle-based algorithms where optimal structures remain unknown.

\section{Discussion}\label{sec12}

Since the early developments in quantum machine learning—including our own work on quantum recommendation systems~\cite{kerenidis2016quantum}—the field has witnessed both significant enthusiasm and critical scrutiny. Recent studies have examined limitations through dequantization results~\cite{tang2019quantum}, demonstrations of classical simulability for certain quantum neural networks~\cite{bermejo2024qcnn_shadows}, and the identification of barren plateaus in specific parameter regimes~\cite{mcclean2018barren}. While such work is valuable for delineating the scope and challenges of quantum machine learning, it should not be interpreted as conclusive evidence against its future utility. The results presented here, in which quantum intelligent agents autonomously rediscover seminal quantum algorithms, offer concrete demonstrations that these methods can be both effective and interpretable.

Table~\ref{tab:summary} summarizes the quantitative performance of our agents across all four benchmark tasks. In each case, training converged within 300 epochs, with agents achieving near-optimal or optimal performance without prior knowledge of target solutions. These results demonstrate the framework's capability for autonomous algorithmic discovery across diverse quantum information processing tasks.

\begin{table}[t]
\centering
\begin{tabular}{|l|l|c|c|}
\hline
\textbf{Task} & \textbf{Metric} & \textbf{Learned} & \textbf{Optimal} \\ \hline
QFT ($n=4$) & Fidelity & 0.999999 & 1.000000 \\ \hline
QFT ($n=6$) & Fidelity & 0.999999 & 1.000000 \\ \hline
Coin Flip (Alice) & $P^*$ & 0.749992 & 0.750000 \\ \hline
Coin Flip (Bob) & $P^*$ & 0.749985 & 0.750000 \\ \hline
CHSH & $F$ & 0.853553 & 0.853553 \\ \hline
Conflicting-Interest & $F$ & 0.640154 & 0.640165 \\ \hline
Grover ($N=4$, 1 query) & $P_{\text{success}}$ & 0.999999 & 1.000000 \\ \hline
Grover ($N=8$, 1 query) & $P_{\text{success}}$ & 0.781249 & 0.781250 \\ \hline
Grover ($N=8$, 2 queries) & $P_{\text{success}}$ & 0.944770 & 0.945313 \\ \hline
\end{tabular}
\caption{Summary of learned agent performance across all benchmark tasks. Agents consistently match or approach optimal values without prior knowledge of target solutions.}
\label{tab:summary}
\end{table}

An important observation is that the discovery of new quantum algorithms in our framework fundamentally relies on access to a quantum computer or a quantum simulator during training. Classical neural networks without such access cannot, in general, evaluate the performance of candidate quantum circuits for inherently quantum tasks, because computing the reward itself entails simulating quantum evolution and measurements. In other words, a purely classical agent without explicit modelling of quantum mechanics cannot feasibly "play" quantum strategies or compute their payoffs. This highlights that our quantum intelligent agents do not simply replace a classical model with a quantum one; rather, they leverage genuine quantum execution during training, making them uniquely positioned to explore algorithmic spaces inaccessible to classical learning systems.

A natural question is that our current experiments use only a few qubits and therefore might not require a quantum computer. We agree that for mathematically self-contained tasks with meaningful small instances (e.g., QFT-type subroutines, nonlocal games, or oracle models), few-qubit settings—or classical simulators—suffice to test whether an agent uncovers the correct algorithmic structure. Our proof-of-concept results operate in this regime for clarity and reproducibility. Crucially, however, whether quantum hardware is \emph{necessary} is problem-dependent and often dictated by input encoding rather than by the learning framework. Many practically relevant tasks exhibit a large \emph{minimum} instance size: for example, binary portfolio selection over $N$ assets already requires $N$ decision qubits (plus ancillas for budget and risk constraints and device-level overhead), so even modest formulations naturally push into the hundreds of qubits. Likewise, searching for hardware-tailored error-correction strategies on devices with hundreds to thousands of qubits inherently lies beyond classical simulability; in such regimes, reward evaluation for candidate quantum policies requires quantum execution.

Our findings open several promising avenues for further research. A central long-term objective is the discovery of genuinely novel quantum algorithms and protocols through the aid of quantum agents. Potential targets include generalizations of the Fourier transform, improved quantum error-correcting codes, and novel cryptographic primitives. Quantum agents could also be deployed in competitive or cooperative environments—either in classical or quantum games—or as benchmarking tools for quantum hardware by having them compete or collaborate in well-defined tasks.

A particularly compelling application domain is quantitative finance. Trading and portfolio management can be naturally formulated as sequential decision-making problems under uncertainty, a setting where reinforcement learning has already demonstrated significant impact in classical contexts. Within our framework, agents parameterized by quantum circuits and trained through interaction with stochastic market simulators could be adapted to autonomously discover trading strategies, hedging policies, or portfolio rebalancing protocols. This vision aligns with recent work applying quantum machine learning to financial forecasting and risk management \cite{Cherrat2023,Thakkar2024}. By framing markets as multi-round games with uncertain payoffs, quantum agents may extend their role in algorithmic discovery toward the development of adaptive financial strategies. While deployment remains a long-term prospect, such applications highlight the broader potential of quantum intelligent agents beyond discovery of quantum algorithms and protocols.

An additional opportunity lies in hardware co-design. The agent-based search for optimized quantum algorithms could guide the development of application-specific quantum processors with fixed circuit architectures, rather than aiming exclusively for universal quantum computers. For example, if agents repeatedly converge to certain fixed-depth, structured circuits for a given task, these could inform the design of specialized processors with tailored connectivity and native gates. Such co-design could reduce compilation overhead, simplify control requirements, and accelerate the practical deployment of quantum technologies.

We are at the outset of what we term the quantum intelligent era, in which the integration of quantum computing and artificial intelligence will not only push the boundaries of algorithmic innovation but may also inform the design of the next generation of quantum devices. The synergy between these fields has the potential to redefine computational limits, advance theoretical understanding, and unlock practical applications. Realizing this vision will require sustained collaboration between experts in quantum algorithms, AI, and hardware engineering—a challenge the community is well positioned to embrace.

\bibliographystyle{unsrtnat}
\bibliography{paper} 
\newpage
\appendix
\section{Trained Quantum Circuits for the Coin Flipping Protocol}

Here we give the two trained quantum circuits for the coin flipping protocol: one for the scenario where Alice is cheating, and one for the scenario where Bob is cheating.

\subsection{Cheating Alice Circuit}


\subsection{Cheating Bob Circuit}
%

\end{document}